\documentclass[twoside,twocolumn,english,aps,showpacs,prb,superscriptaddress]{revtex4-2}
\pdfoutput=1 
\usepackage[T1]{fontenc}
\usepackage{color}
\usepackage{bm}
\usepackage{amsmath}
\usepackage{amssymb}
\usepackage{graphicx}
\usepackage{esint}
\usepackage{mathrsfs}
\usepackage{booktabs}
\usepackage{multirow}
\usepackage{makecell}
\usepackage{verbatim}
\usepackage[plainpages = false, pdfpagelabels, 
                 bookmarks,
                 bookmarksopen = true,
                 bookmarksnumbered = true,
                 breaklinks = true,
                 linktocpage,
                 colorlinks = true,
                 linkcolor = blue,
                 urlcolor  = blue,
                 citecolor = blue,
                 anchorcolor = green,
                 hyperindex = true,
                 hyperfigures
                 ]{hyperref} 
\usepackage{dcolumn}   

\makeatletter
\newcommand*{\rom}[1]{\expandafter\@slowromancap\romannumeral #1@}
\makeatother

\usepackage{times}{\Large }

\makeatother

\usepackage{babel}

\begin{document}

\title{Enhanced localization in the prethermal regime of continuously measured many-body localized systems}

\author{Kristian Patrick}
\email[]{kpatrick@baqis.ac.cn}
\affiliation{Beijing Academy of Quantum Information Sciences, Beijing 100193, China}
\author{Qinghong Yang}
\affiliation{State Key Laboratory of Low Dimensional Quantum Physics, Department of Physics, Tsinghua University, Beijing, 100084, China}
\author{Dong E. Liu}
\email[]{dongeliu@mail.tsinghua.edu.cn}
\affiliation{State Key Laboratory of Low Dimensional Quantum Physics, Department of Physics, Tsinghua University, Beijing, 100084, China}
\affiliation{Beijing Academy of Quantum Information Sciences, Beijing 100193, China}
\affiliation{Frontier Science Center for Quantum Information, Beijing 100084, China}
\affiliation{Hefei National Laboratory, Hefei 230088, China}

\begin{abstract}
Many-body localized systems exhibit a unique characteristic of avoiding thermalization, primarily attributed to the presence of a local disorder potential in the Hamiltonian. In recent years there has been an interest in simulating these systems on quantum devices. However, actual quantum devices are subject to unavoidable decoherence that can be modeled as coupling to a bath or continuous measurements. The quantum Zeno effect is also known to inhibit thermalization in a quantum system, where repeated measurements suppress transport. In this work we study the interplay of many-body localization and the many-body quantum Zeno effect. In a prethermal regime, we find that signatures of many-body localization are enhanced when the system is coupled to a bath that contains measurements of local fermion population, subject to the appropriate choice of system and bath parameters. 
\end{abstract}

\maketitle

\section{Introduction}
We are currently living in the noisy-intermediate-scale-quantum (NISQ) era of quantum computing, meaning that devices are subject to noise that will influence the outcome of computations\citep{preskill2018quantum}. This noise enters as an undesired coupling to an external environment due to our limitations in engineering perfect (closed) quantum systems. Following recent progress, namely the simulation of many-body localization (MBL) on a quantum simulator~\citep{smith2016many,xu2018emulating,guo2018detecting,zhu2021probing}, one may ask how imperfections in device engineering may affect the outcome of the simulation. We expect, in general, that the presence of an environment will destroy localization at long times. This is supported by recent theoretical investigations into open system MBL~\citep{znidaric2,nandkishore2014spectral,levi2016robustness}. However, the full extent to which an environment influences the system is still an open question. This is exemplified in Ref.~\cite{nandkishore2015many}, where it is shown that an appropriately modeled bath may lead to survival of the localized phase. 

Recently, the impact of repeated measurements on quantum systems has garnered significant attention, particularly concerning measurement-induced criticality~\cite{li2018quantum,chan2019unitary,skinner2019measurement,hoke2023quantum,poboiko,poboiko2,popperl}. This phenomenon occurs when measurements within a random quantum circuit~\cite{nahum2017quantum,khemani2018operator,sunderhauf2018localisation} prompt an entanglement transition from volume law to area law, contingent on a sufficiently high measurement rate~\cite{jian2020measurement}. The regime of area law growth of entanglement is known as the Zeno limit, due to its analogy to the quantum Zeno effect~\cite{misra1977zeno} where repeated measurements inhibit transport in the system of interest. In exploring the many-body quantum Zeno effect it has been found that the action of measurements can effectively induce a slow-down effect in the system~\cite{Huse2015Localized,maimbourg2021bath,ippoliti2021entanglement,halimeh2022enhancing}, by disentangling the system and suppressing transport. In addition, an effective field theory, termed the Keldysh nonlinear sigma model, has been developed and analytically shows that the measurements induce a slowdown or diffusive behavior~\cite{yang2023keldysh}.

In this work we aim to expose a new phenomenon in open system MBL: a prethermalization regime with enhanced localization. 
There exists a significant and ongoing debate within the scientific community regarding the existence and characteristics of a true MBL phase in infinitely large systems~\cite{suntajs2020quantum,abanin2021distinguishing, sierant2022challenges,long2023pheno}. Here, we specifically focus on the prethermal regime or finite-time MBL that is applicable to current NISQ devices, where we expect signatures of MBL to be robust. Thus, the prethermal regime we discuss here is an artifact of open system dynamics and not of ``finite-size MBL''.
In order to expose the phenomenon of prethermal enhancement, our approach involves selecting a bath characterized by its continuous measurement or projection onto occupied sites. This methodological choice gives rise to two conflicting effects. The first involves localization, which arises from the system's coherent evolution in response to static disorder~\citep{oganesyan2007localisation,pal2010many,serbyn2013universal,alet2018many,abanin2019colloquium,popperl2}. The second concerns the preservation of information, attributable to the many-body quantum Zeno effect due to continuous measurement. For the noninteracting case, the interplay of disorder and measurements has been studied for free-fermion systems, for example in Refs.~\cite{Szyniszewski2023disordered,Szyniszewski2024unscrambling}, where it is shown that measurements may destroy an Anderson localized phase or disorder may stabilize the critical phase of a monitored system. However, this work extends these studies to include interactions, which may expose behaviors beyond those in the free-fermion setting. 

Our numerical results show that when the effective rate of measurement is tuned to an appropriate value, the measurements will enhance the localization for a particular time window during the coherent evolution.
The numerical results can be qualitatively elucidated through analytic discussion. To exemplify our result we study the dynamics of an archetypal MBL system~\cite{serbyn2013universal}, a $1$D chain of interacting fermions with static disorder. We choose the bath to be Markovian and represented by local fermion number operators $n_j$. This choice of bath preserves the particle number of the system and is closely related to experimental setups. For example, in Ref.~\cite{sarkar2014light} it is shown that a fermionic optical lattice quantum simulator is subject to decoherence of this form due to light scattering. 

\subsection{Outline of results} 
Assuming Markovian dynamics, behaviour of an open quantum system is dictated by solutions to the Lindblad master equation $\mathcal{L}\rho(t)=\partial_t \rho$ given by ($\hbar=1$)
\begin{align}\label{eq:lme}
\mathcal{L}\rho(t) = -i\left[H,\rho\right] + \gamma\sum_j L_j\rho L_j^\dagger -\frac{1}{2}\lbrace L_j^\dagger L_j,\rho\rbrace,
\end{align}
where $\mathcal{L}$ is the Lindblad superoperator.  By selecting local fermionic number operators as the set of measurement operators, $L_j=n_j$, we identify three regimes of interest that depend on the strength of the coupling to the bath, $\gamma$. This can also be viewed as an effective rate for which measurements are made on the system. First, we note that for any $\gamma\neq0$ we expect a thermal steady state. To see this, consider steady state solutions $\partial_t \rho_\infty =0$ to the Lindblad master equation when the Lindblad operators $L_j$ satisfy $L_j^2=L_j=L_j^\dagger$. It can be easily shown that $\rho_\infty=\mathbb{I}$ is a solution to Eq.~(\ref{eq:lme}) under these conditions. So, in the long time limit we expect all states to thermalize under the Lindblad master equation. 

In the first regime, for very small $\gamma$, the behavior of the system will be only weakly affected by the measurements. 
Thus, the system largely maintains its localization attributes akin to the closed system over an extended duration, yet ultimately thermalizes. Measurements gradually dismantle system localization by randomly projecting a site onto a definitive occupation state, which may not align with initially occupied sites. Over time, this leads to equal occupation across all sites, culminating in a thermal-appearing state.

In the second regime, with very large $\gamma$, we expect the effect of the measurements to dominate the behavior of the system. This regime has been studied in detail recently~\cite{li2018quantum,skinner2019measurement,fuji2020measurement,sang2021entanglement}, under the guise of measurement induced criticality. For sufficiently strong measurement rate the system enters a disentangling phase where entanglement is suppressed, but the result is that excitations are evenly distributed throughout the system. 

The third and final regime is for an intermediate coupling to the bath, where we see a prethermalization regime that has an enhancement in localization. Here, the interplay between unitary evolution and projective measurements results in a temporary stability with an enhanced localization signature, before an eventual thermalization. The temporary enhancement results from occupied sites being the most probable to be measured in the trajectory approach. For a large enough rate of measurement this preserves more information of the initial state for some time window, without completely dominating the dynamics. 

\section{Interpreting enhancement from the similarity between measurements and disorders} Note that the Lindblad master equation Eq. \eqref{eq:lme} with jump operator $L_j=n_j$ for the unconditional measurements also describes the effect of dephasing noise. Existing works reveal slowdown and heating effects of such noise~\cite{znidaric,znidaric2,sieberer,jin,turkeshi}, which will even destroy many-body localization~\cite{medvedyeva,wu,luschen,yamamoto2023localization}.  Therefore, the prethermal enhancement of localization observed in this work is rather surprising. In recent work~\cite{yang2023keldysh}, it is shown that free fermions subjected to continuous measurements (also referred to as ``monitored systems'') can be accurately described by a special Keldysh nonlinear $\sigma$ model. Consequently, this framework unifies monitored and disordered fermions~\cite{finkelstein} within the Keldysh field theory. Specifically, the corresponding Keldysh action of Eq.~\eqref{eq:lme} is in a similar form to that of a disordered free-fermion gas~\cite{finkelstein,horbach,kamenev}, and in the long-time steady regime, a Drude-form conductivity is obtained. In addition to the similarity, the measurement case will introduce an extra decoherence effect due to the nature of open systems. This decoherence effect will turn the quantum system into a classical one, and then hides similar exotic effects as that produced by disorders in a quantum system, such as many-body localization.

\begin{figure*}[t!]
    \centering
	\includegraphics[width=2\columnwidth]{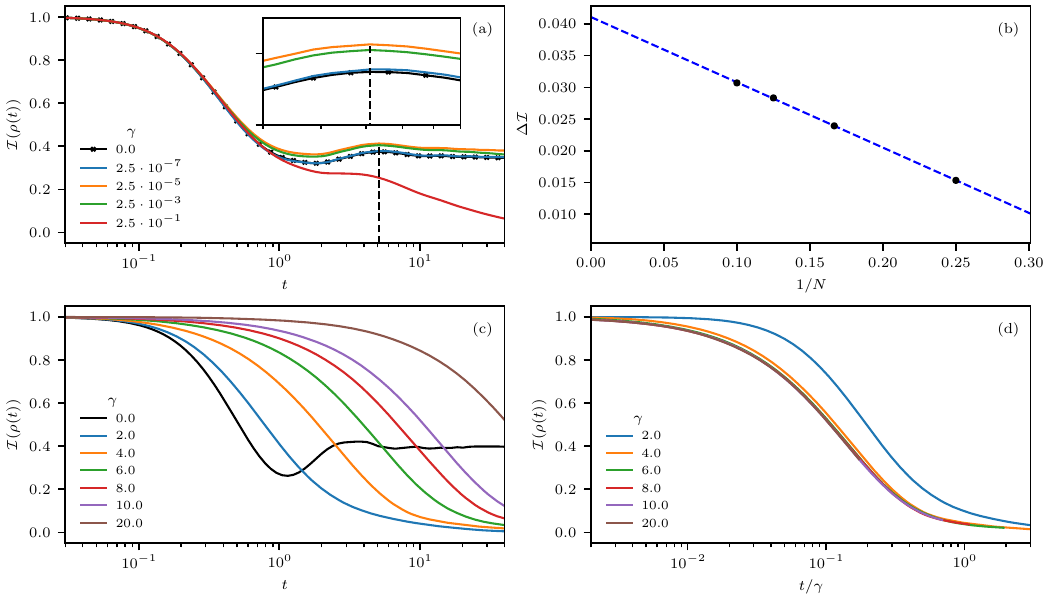}
	\caption{
		(a) Time evolution of imbalance for a chain of $N=10$ sites, $V=2$, and $W=5$, with a bath consisting of $L_j=n_j$ measurements, where $\gamma$ spans many orders of magnitude. We see that within some window of $\gamma$ there is an enhancement in localization. All data points are averaged over $n_d=300$ disorder realizations. The inset zooms in to expose the enhancement clearly. (b) Finite-size scaling of the enhancement $\Delta\mathcal{I}$, at the value of $t$ indicated by the vertical dashed line in (a) and between $\gamma=0$ and $\gamma=2.5\cdot 10^{-3}$. System sizes $N=4,6,8$ and $10$ are averaged over $n_d = 10^3$, $1500$, $600$, and $300$ realizations, respectively. (c) Time evolution of imbalance for a chain of $N=6$ sites, $V=2$ and $W=5$, where $\gamma$ now spans values deep into the large measurement regime. (d) The same data as in (c), but with the time axis rescaled by $1/\gamma$, showing universal behavior when $\gamma>2$.}
	\label{fig:imbalance}
\end{figure*}

From this discussion, one finds that the similarity with disorders can be further revealed when the decoherence effect is suppressed or not fully established in the continuous measurement case. A possible way to reduce the decoherence effect is to consider small values of measurement strength $\gamma$ and short-time dynamics of the monitored system. In this scenario, the decoherence  is small at each time step and will not sufficiently accumulate in a short time period. Indeed, this is the regime at which we observe the enhancement of many-body localization shown in Figs.~\ref{fig:imbalance}--\ref{fig:correlations}. In this regime, decoherence is not fully established and measurements will act like disorders in a quantum system; then the disorder strength $W$ is in turn effectively increased, and thus the enhancement of many-body localization appears. Our study provides numerical evidence demonstrating that a unique signature observed in disordered systems is also present in systems under continuous measurement, thereby further illustrating the applicability of the corresponding field theory framework~\cite{yang2023keldysh}.

\section{Model Setup} 
\subsection{Hamiltonian}
We choose the system to consist of spinless fermions hopping on a $1$D lattice of $N$ sites, subject to nearest-neighbor interactions and random onsite disorder. The Hamiltonian is
\begin{align}
\!\!\! H=\sum_j^{N-1} -(c_{j}^\dagger c_{j+1} + c_{j+1}^\dagger c_{j}) + Vn_jn_{j+1} + \sum_j^Nw_jn_j,
\end{align}
with $c_j^\dagger(c_j)$ the fermionic creation(annihilation) operator at site $j\in{1,\dots,N}$, $n_j$ the local fermionic number operator, $V$ the nearest-neighbor interaction strength and $w_j\in[-W,W]$ a random onsite disorder taken from a uniform distribution. Throughout the majority of this work, the initial state is chosen to be the half-filled N\'eel state $\left|\psi(0)\right\rangle=\left|01\dots 01\right\rangle$, which can be prepared in superconducting circuits~\cite{xu2018emulating} and optical lattice~\cite{schreiber2015observation,guo2018detecting} setups. This model has a known MBL phase transition and has been extensively studied in a variety of settings~\cite{pal2010many,serbyn2013universal,luitz2015many,fischer2016dynamics,levi2016robustness,everest2017role,hamazaki2019non}. The Lindblad operators are chosen to be measurement of local fermion number, $L_j\in\{n_j\}_{j=1}^N$, with coupling strength $\gamma$~\cite{levi2016robustness,everest2017role,fuji2020measurement}. The numerical implementation of the dynamics is provided and well maintained by the Quantum Toolkit in Python (QuTiP) package~\citep{JOHANSSON20121760}.

\subsection{Diagnostic tools} 
In order to analyze the localization properties of this system we study the imbalance, measuring the distribution of fermion density along the chain. We employ the generalized imbalance definition from~\cite{guo2021observation}, which can be utilized for any configuration of initial state. The imbalance operator is defined as $I=\sum_j\beta_j n_j$ with $\beta_j=\frac{1}{N_1}(-\frac{1}{N_0})$ for a qubit initialized in the state $\left|1\right\rangle(\left|0\right\rangle)$. Then, the imbalance can be defined simply as 
\begin{align}\label{eq:imbalance}
    \mathcal{I}(\rho(t))=\mathrm{Tr}[I\rho(t)].
\end{align}
For the initial state given above we have $\mathcal{I}(\rho(0))=1$. For a thermalized steady state fermions will be evenly distributed along the chain resulting in $\lim_{t\to\infty}\mathcal{I}(\rho(t))=0$.

To accurately capture the quantum correlations in the system we study the logarithmic negativity, an entanglement measure for mixed states~\cite{vidal2002computable,plenio2005logarithmic,calabrese2012entanglement,sang2021entanglement}. It is defined as the trace norm of the partially transposed density matrix, 
\begin{align}\label{eq:log_neg}
S_n(t)= \ln ||\rho^{T_A}||,
\end{align}
where $||O||=\mathrm{tr}|O|=\mathrm{tr}\sqrt{O^\dagger O}$ is the trace norm of operator $O$ and $\rho^{T_A}$ is the partial transpose of the subsystem $A$ computed for fermions, as described in~\cite{shapourian2018entanglement}. Note that throughout this study all bipartitions have equally weighted subsystems with $N_A=\frac{N}{2}$. To support the logarithmic negativity we also study correlations between sites of the chain via a $2$-point correlation function $\left\langle c_j^\dagger c_k\right\rangle=\mathrm{Tr}\rho  c_j^\dagger c_k$. 

As a final diagnostic tool we study the von Neumann entropy of the mixed state subsystem density matrix, given by 
\begin{align}\label{eq:vne}
S(t)=-\mathrm{Tr}\rho_A \ln \rho_A,
\end{align}
where $\rho_A$ is the reduced density matrix of the mixed state $\rho$. Unlike the logarithmic negativity, the von Neumann entropy will capture both quantum and classical correlations. It is therefore not a good measure of entanglement, but it remains a useful tool to diagnose nontrivial behavior and the build up of correlations in the system. 

 \begin{figure}[t!]
    \centering
	\includegraphics[width=1\columnwidth]{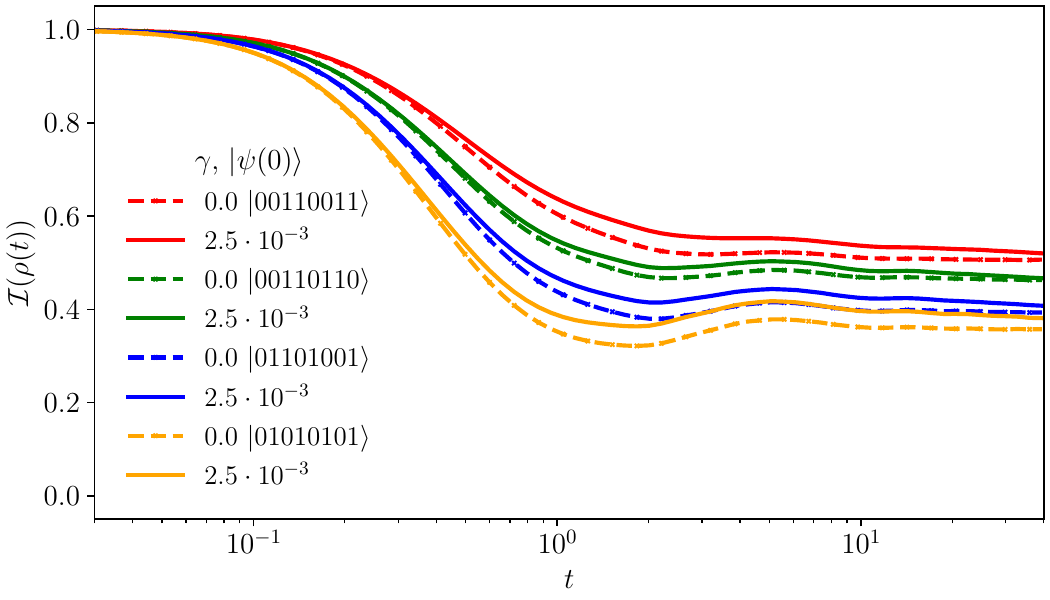}
	\caption{For a chain of $N=8$ sites, $V=2$, and $W=5$, we show the imbalance for a variety of different initial conditions over $600$ distinct disorder realizations. Dashed lines show the evolution of imbalance without influence of the bath $\gamma=0$; solid lines show the enhanced imbalance for a given $\gamma\neq 0$. Initial states are given in the legend alongside the respective $\gamma=0$ legend item.}
	\label{fig:transient}
\end{figure}

\section{Simulation Results}
\subsection{Imbalance} 
The main result can be exemplified in Fig.~\ref{fig:imbalance} (a), where the time evolution of the imbalance is shown for $L_j=n_j$ as the Lindblad operators. When $\gamma=0$ the imbalance reaches a steady state value that signals localization in the system. Note that in order to expose dominant features in the following numerical simulations we employ a moving average routine~\cite{Hyndman2011} to suppress noise and oscillatory behavior. For transparency, in Appendix~\ref{sec:correction} we present the same data without employing a moving average routine. 

 \begin{figure*}[t!]
    \centering
	\includegraphics[width=2\columnwidth]{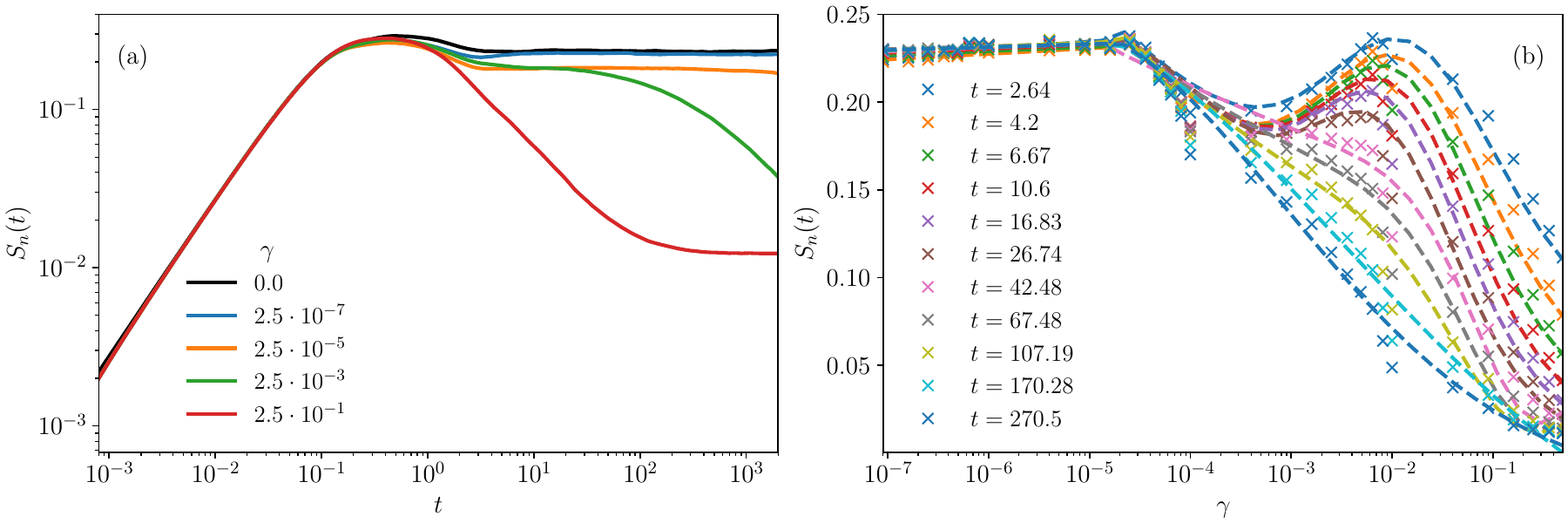}
	\caption{For a chain of $N=6$ sites, $V=2$, and $W=5$, we show (a) the logarithmic negativity against time after a partial transpose of half the system for various choices of $\gamma$, and (b) the logarithmic negativity against $\gamma$ for many different time slices. In (a) we see that there is a reduction in entanglement as the influence of the bath becomes more dominant.}
	\label{fig:entanglement}
\end{figure*}

For very small $\gamma$, such that the effective rate of measurement occurs a long time after the closed system reaches a steady state, we see that the prethermal regime is dictated by the coherent dynamics alone. After sufficient time has passed $\sim 1/\gamma$ the system begins to lose information and reduces to an $\mathcal{I}=0$ thermal state. For intermediate $\gamma$, where the rate of measurement occurs at a similar time to the closed system reaching a steady state, we see an enhancement in localization for a time window before an eventual thermalization. We thus call this a prethermalization regime with enhanced localization. This regime is highlighted in the inset of Fig.~\ref{fig:imbalance} (a), where the region of interest is magnified. In Fig.~\ref{fig:imbalance} (b) we show a finite-size scaling of the enhancement $\Delta\mathcal{I}(\tau)=\mathcal{I}(\rho_{\gamma_1}(\tau))-\mathcal{I}(\rho_{\gamma_0}(\tau))$, for $\gamma_0=0$, $\gamma_1=2.5\cdot10^{-3}$, and $\tau=4.64$. There is a clear increase in $\Delta\mathcal{I}$ as we increase system size, indicating that the enhancement is robust for large system sizes and is not only a small system size effect. In Fig.~\ref{fig:imbalance} (c), we see how large $\gamma$ influences $\mathcal{I}$. Here, information is preserved for a longer time, but with an eventual thermalization without an intermediate prethermalization regime characteristic of measurement-induced localization. Figure~\ref{fig:imbalance} (d) shows the same data but with the time axis rescaled by $1/\gamma$. This exposes universal behavior in the large measurement regime, when $\gamma\gtrsim V$, and indicates that all coherent dynamics attributed to MBL have been lost. For the remainder of this work we focus on the regime prior to this, where there is an interplay between MBL dynamics and continuous measurements.

\subsubsection{Prethermal behaviour}

For a system to be established as having prethermal behavior, it is required to identify separate timescales dominating the dynamics. Following the definition in~\cite{mori2018thermalization}, we draw analogy from closed quantum systems where a Hamiltonian may be written as $H=H_0+\lambda V$, with $\lambda$ a dimensionless parameter. For short timescales the dynamics are dominated by $H_0$ and the system will equilibrate to a prethermal state dictated by $H_0$. At longer timescales, when the effect of $\lambda V$ becomes relevant, the system will equilibrate to the true steady state of $H$. In this work, the two separate timescales are dictated by unitary dynamics and measurements by the environment.

To conclusively establish the behavior observed in Fig.~\ref{fig:imbalance} as prethermal and initial state independent, we show in Fig.~\ref{fig:transient} the enhancement for a number of different random initial configurations that are in the half-filled fermion number sector. To select these initial states, we first identify all possible product state configurations in the half-filled sector and associate them with an index. We then use a random number generator to select an index, ensuring previously selected initial states are removed from the pool. From Fig.~\ref{fig:transient} it is clear that regardless of the choice of initial state, we still identify an enhancement in localization. The enhancement manifests as an interplay of timescales, where the effects of the analogous $\lambda V$ become relevant at a similar timescale dictated by $H_0$.

\subsection{Logarithmic negativity} 
In Fig.~\ref{fig:entanglement} (a), we simulate the evolution of entanglement via the logarithmic negativity $S_n(t)$. We see an initial linear growth in entanglement that overshoots and plateaus off when $\gamma=0$. This cutoff is dictated by the system size and we expect it to grow logarithmically as $N$ is increased~\cite{abanin2019colloquium}. For very small $\gamma$ we see very little deviation in the plateau, as expected from the analysis of imbalance in Fig.~\ref{fig:imbalance}. For values of $\gamma$ where there is an enhancement of localization, we see a reduction in entanglement, evidenced by a plateau at a smaller value of $S_n(t)$. This is in line with the intuition that when more initial state information is retained there is less build up of correlations and therefore less entanglement between subsystems. For very large $\gamma$ the dynamics are entirely dominated by the disentangling nature of the environment. 

In Fig.~\ref{fig:entanglement} (b), we show how $S_n(t)$ changes with increasing $\gamma$ for different choices of time. For small values of $\gamma$ we see very little deviation in $S_n(t)$, signaling that the bath does not influence dynamics at this timescale. At intermediate $\gamma$ there is a clear dip in $S_n(t)$ that corresponds to the regions in Fig.~\ref{fig:imbalance} for which we see an enhancement in localization. For larger $\gamma$ there is a sharp decrease in the entanglement between subsystems, dominated by the influence of the environment. 

\begin{figure*}[t!]
    \centering
	\includegraphics[width=2\columnwidth]{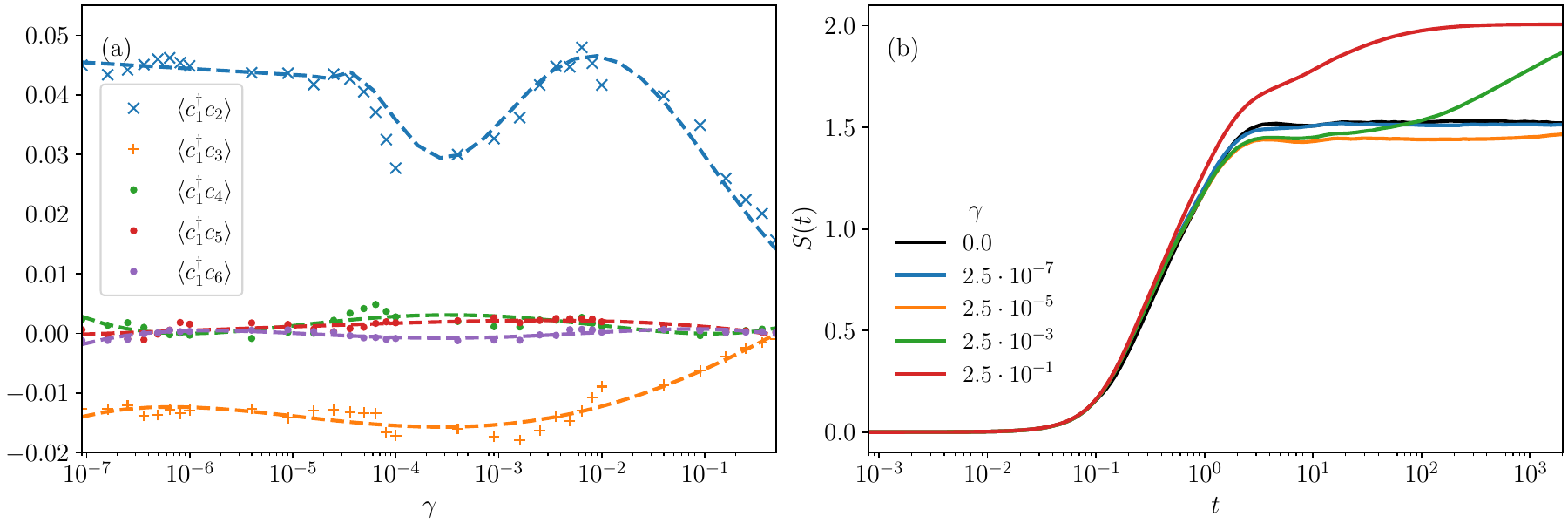}
	\caption{For a chain of $N=6$ sites, $V=2$, and $W=5$, we show (a) two-point correlations from the first site to all other sites across many values of $\gamma$ at time $t=10.6$s, and (b) the change in von Neumann entropy with time for various choices of $\gamma$. In (a), there is a signature of reduced entanglement in the nearest-neighbor correlation between sites $1$ and $2$ for an intermediate region of $\gamma$ that is in agreement with (b). }
	\label{fig:correlations}
\end{figure*}

\subsection{Correlations} 
In Fig.~\ref{fig:correlations} (a), we show $2$-point correlations in the chain for varying coupling strength at time $t=10.6s$. Most enlightening is the nearest-neighbor correlation. Initially this remains close to constant and is fitted with a linear function. After reaching large enough $\gamma$ we see a reduction in its values, signaling less information transported along the chain at this time. The eventual rise in value signals a departure from the enhanced regime, before a second decay as the system enters the Zeno phase. The region of the dip is in exact correspondence with the dip seen in Fig.~\ref{fig:entanglement} (b) that determines the region of enhanced localization. 

Finally, in Fig.~\ref{fig:correlations} (b) we show how the von Neumann entropy changes with time for different choices of $\gamma$. Keeping in mind that this includes both quantum and classical correlations, we now see that for large $\gamma$ the value of $S(t)$ increases beyond the $\gamma=0$ case, even though $S_n(t)$ decreases.  
For very small $\gamma$, at the timescale shown, the system behaves much like the closed system as expected from previous figures. But in the intermediate regime there is an initial plateau with a lower entropy than for $\gamma=0$, before an eventual growth. This can be attributed to a reduction in quantum correlations due to the enhanced localization, followed by a growth in thermalization or classical correlations dictated by the environment. 
While this measure is not a true quantifier of quantum correlations, it does help distinguish subtle differences in the dynamics of an open quantum system and aids in building a more intuitive understanding of the system dynamics.

From these simulations we can conclude that there exists a finite window where localization is (atleast temporarily) enhanced by the presence of the environment. For very small values of coupling there is little effect on the system. For very large coupling the environment dominates the behavior and the system disentangles. While this results in a slowdown in the thermalization process, as evidenced by the imbalance and slow growth of entropy, it will always lead to thermalization in the system. 

\section{Lieb-Robinson Bound Analysis}
\subsection{Lieb-Robinson bound and prethermal regime} 
The results presented above can be supported by considering the speed at which information propagates along the chain~\cite{burrell2008information,barthel2012quasilocality,kim2014local,deng2017logarithmic,vu2023dissipative}. Here, we utilize the results in Ref.~\cite{burrell2008information} to obtain a Lieb-Robinson bound for an open system influenced by measurements satisfying $L_j\equiv P_j=P_j^2$. The Lieb-Robinson bound can be expressed as an upper bound on the Lieb-Robinson commutator~\cite{lieb1972the,burrell2007bounds,burrell2008information}
\begin{align}
C_B(x,t) = \sup_{A_x} \frac{\left|\left|\left[A_x,B(t)\right]\right|\right|}{\left|\left|A_x\right|\right|},
\end{align}
where $\left|\left|O\right|\right|$ is the operator norm of the operator $O$, $A_x$ is an operator acting on site $x$ and $B(t)=\mathbb{E}_\xi B_\xi(t)$ is an operator in the Heisenberg picture that may have support over many sites, disorder-averaged over realizations $\xi$. To determine $B(t)$ we require the adjoint Lindblad master equation to be given by $\mathcal{L}^\dagger B(t) = i\left[H,B(t)\right] -\gamma\mathcal{D}(B)$ where $\mathcal{D}(B(t))$ is the dissipator, the term in Eq.~(\ref{eq:lme}) containing the Lindblad operators. 
 
Following the steps taken in~\cite{burrell2008information}, and with full details of the mathematical steps in Appendix~\ref{sec:lrb}, we find the following bound for an open system under the influence of measurements,
\begin{align}\label{eq:lrb_pm}
C_B(x,t)\leq e^{(-2\gamma p+4\left|\left|H\right|\right|)t}\sum_j^Ne^{2\left|\left|H\right|\right|R_{x,j}t}C_B(j,0),
\end{align}
where we assume $\sum_jP_jB(t)\leq pB(t)$ and $R_{j,k}=\delta_{j,k+1}+\delta_{j+1,k}$. Assuming no particle loss in the system, we may choose our measurement operator to be the fermionic occupation number, i.e., $P_j=n_j$, so that $p\equiv\sum_j n_j$ the total number of fermions in the system. In our work we initialize the system at half filling, so $p=N/2$, and this is fixed throughout the dynamics.

Given that $\left|\left|R\right|\right|=2$, if $\gamma>\frac{4\left|\left|H\right|\right|}{p}$ then the exponentially decreasing part of our bound is dominant. When $t<1/\gamma$ the open-system dynamics remain negligible and dynamics will follow the closed chain. When $t\sim 1/\gamma$ the closed-system bound~\cite{kim2014local} is no longer applicable and we must consider the dynamics induced within the constraint of the dominant part of our bound. This bound is negligible when either $|j-k|\geq tc_\chi$ for a constant $c_\chi >0$ or when $t\geq t_\chi= \log(\frac{C}{\chi})/\Gamma$, with $C=\sum_jC_B(j,0)$, $\Gamma=2\gamma p+8\left|\left|H\right|\right|$, and $\chi$ a constant. Here, $\Gamma$ is chosen to cancel any contribution from the exponentially growing part of the bound. Combining these two conditions, we have that the bound is non-negligible only when $\left|j-k\right|<c_\chi t_\chi$, which is a constant. This constant implies localization as correlations can only spread up to a finite number of sites. However, for very long times and small system sizes, the condition $|j-k|\geq tc_\chi$ implies that correlations may still spread across the entire system. In this case we expect an eventual thermalization that follows a prethermalization regime. 

\subsection{Lieb-Robinson bound outside the prethermal regime} 
To quantify the applicability of Eq.~(\ref{eq:lrb_pm}) for values of $\gamma$ outside the prethermal regime, it is important to reference other bounds related to this system. First, for a closed MBL system with interactions and disorder there exists a Lieb-Robinson bound that results in a logarithmic light cone~\cite{kim2014local,deng2017logarithmic}. Second, for general open quantum systems with Markovian dynamics the Lieb-Robinson bound has been shown to have a ballistic light cone for Liouvillians that are the sum of local parts~\citep{barthel2012quasilocality}. In both cases, outside of the light cone correlations decay exponentially fast. The main difference between Eq.~(\ref{eq:lrb_pm}) and the general bound for open quantum systems is that the general bound will always allow for ballistic spreading of correlations, causing the system to thermalize. On the other hand, if the exponentially decaying part of Eq.~(\ref{eq:lrb_pm}) is dominant then it is possible for information to be retained for some finite time when the system undergoes time evolution. In the limit that $\gamma\to 0$, Eq.~(\ref{eq:lrb_pm}) returns to a bound where correlations spread ballistically. However, it should be noted that Eq.~(\ref{eq:lrb_pm}) does not consider any microscopic details of the Hamiltonian, so there may exist a tighter bound when taking these details into account.

Returning to the bound derived in Eq.~(\ref{eq:lrb_pm}), and keeping in mind that $\left|\left|R\right|\right|=2$, if $\gamma<\frac{4\left|\left|H\right|\right|}{p}$ then Eq.~(\ref{eq:lrb_pm}) is exponentially growing and the bounds discussed previously will provide a more appropriate restriction on the spread of correlations, which depends on how much time has passed in the evolution. Providing $t<1/\gamma$, the open-system dynamics are negligible and the system will follow the closed-system dynamics with correlations spreading logarithmically with time~\cite{kim2014local}. After sufficiently long time $t\sim 1/\gamma$, the open-system dynamics become relevant and the general open-system bound is applicable resulting in ballistic spreading of correlations. Thus, for very small $\gamma$ we see initial localization properties akin to the closed system, followed by an eventual thermalization. Thus, in the limit $\gamma<\frac{4\left|\left|H\right|\right|}{p}$ we expect a prethermal regime that is dictated by the coherent dynamics alone. 

Conversely, if $\gamma\gg\frac{4\left|\left|H\right|\right|}{p}$, so that the timescale for unitary closed-system dynamics is longer than the effective rate of measurement, the system will only follow the closed-system dynamics for a very short time. Now, similarly to the analysis in the prethermal regime, the bound is negligible when either $|j-k|\geq tc_\chi$ for a constant $c_\chi >0$ or when $t\geq t_\chi= \log(\frac{C}{\chi})/\Gamma$, with $C=\sum_jC_B(j,0)$, $\Gamma=2\gamma p+8\left|\left|H\right|\right|$, and $\chi$ a constant. Combining these two conditions, we have that the bound is non-negligible only when $\left|j-k\right|<c_\chi t_\chi$, which is a constant. This constant implies localization as correlations can only spread up to a finite number of sites. Thus we expect to see localization properties up to a time given by $|j-k|\geq tc_\chi$, after which, due to the finite size of the system, we expect thermalization. This is known as the Zeno regime where frequent repeated measurements retain information in the system.

To exemplify the applicability of this bound, in Fig.~\ref{fig:imbalance}c we considered the case when $\gamma$ is very large so that dynamics are entirely dominated by the bound for both small and large timescales. It is clear that as the value of $\gamma$ increases the amount of time that information of the initial state remains also extends, before an eventual decay to a thermalized state. This is due to the repeated measurements constantly resetting the initially occupied states back to certain occupation. This can also be explained by analyzing the bound. From the conditions above we find that $t_\chi\sim1/\gamma$. Thus, for larger $\gamma$, due to the bound $\left|j-k\right|<c_\chi t_\chi$, the spread of correlations is pinned tighter than for smaller $\gamma$. 

\section{Discussion}
In this work we have exposed a regime of enhanced localization due to the interplay of MBL and Zeno physics. 
When the MBL system is coupled to an external bath, the bath serves as a source of energy and particle exchange, driving the system toward thermalization. However, the Lieb-Robinson bound restricts the speed of information propagation in the system, effectively slowing down the dynamics of the MBL system as it interacts with the bath. Therefore, even though the bath promotes delocalization or thermalization, the process is slowed down by the constraints imposed by Eq.~(\ref{eq:lrb_pm}). The prethermal phase is characterized by the system exhibiting both localized and thermalized features, as the system resists full thermalization for a relatively long time. The details of the prethermal phase depend on factors such as the strength of coupling between the system and the bath, the degree of disorder, and the strength of the interactions in the system. The enhancement of the localization further exemplifies the argument about the unification of projective measurements and disorders~\cite{yang2023keldysh}.

Further, we believe that this result may influence ongoing MBL simulations, for example~\cite{xu2018emulating}, where decoherence effects are ignored due to the dephasing time of qubits being slower than the experimental time. Finally, we envisage that our result will be directly observable on current quantum devices capable of simulating hybrid quantum circuit models, see for example~\cite{li2018quantum}, which consist of both unitary dynamics and random measurements, by interlacing the MBL architecture with random measurements.

\section*{Acknowledgements}
We thank Baiting Liu and Xiao Li for insightful discussions. This work was supported by National Natural Science Foundation of China (Grant No.~92365111), Beijing Natural Science Foundation (Grant No.~Z220002), and the Innovation Program for Quantum Science and Technology (Grant No.~2021ZD0302400).


\setcounter{figure}{0}
\renewcommand{\thefigure}{A\arabic{figure}}

\appendix
\section{Imbalance without moving average smoothing routine}\label{sec:correction}
\begin{figure}[h]
    \centering
	\includegraphics[width=1\columnwidth]{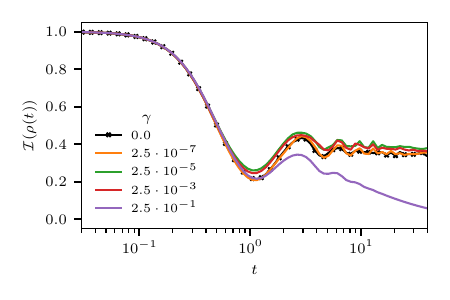}
	\caption{Imbalance for a chain of $N=10$ sites without smoothing using the moving average routine. Even without smoothing, there is clear distinction between the three regimes of interest within the timescales analyzed: MBL dominant for small $\gamma$, enhanced localization for intermediate $\gamma$, and measurement dominant for large $\gamma$.}
	\label{fig:unsmoothed}
\end{figure}

For transparency, in Fig.~\ref{fig:unsmoothed} we show the data presented in Fig.~1 of the main text without using the moving average smoothing routine~\cite{Hyndman2011}. While this routine highlights dominant features, it may also hide less dominant features within the data. It can be seen from the figure that the purpose of using the routine is only to suppress noise fluctuations and to highlight the enhanced localization that is discussed. There remains a clear separation between the curves for no disorder (black) and small disorder (orange) from those with intermediate values of disorder (green and red), with high disorder curves (purple) not showing a prethermal regime. 

\section{Lieb-Robinson bound for an open system under the influence of repeated measurements}\label{sec:lrb}
In this section we present full details for the derivation of a Lieb-Robinson bound for an open system, where the Lindblad operators satisfy $L_j\equiv P_j=P_j^2$. It is based on a similar calculation found in~\cite{burrell2008information}. The following method is general, but makes a number of assumptions. First, assume a time-independent Hamiltonian with nearest-neighbor interactions at most, that is $H=\sum_j h_j$ where $h_j$ acts on sites $j$ and $j+1$. The Hamiltonian contains static disorder, so all observable quantities are calculated as an ensemble average over each disorder realization $\xi$. For example, the density matrix is given by $\rho(t)=\mathbb{E}_\xi \rho(\xi,t)$. Assuming Markovian evolution, we model the dynamics by the Lindblad master equation
\begin{align}
\partial_t \rho(t) &= -i\left[ H,\rho(t)\right]-\gamma\sum_j\left[\left[\rho(t),P_j\right],P_j\right]\\
&=\mathcal{L}\rho(t)\nonumber
\end{align}
where $P_j$ is a measurement operator and $1/\gamma$ is the effective rate of measurement. Note that setting $\gamma\rightarrow\gamma/2$ recovers the standard form of the Lindblad master equation given in the main text. 

A Lieb-Robinson bound can be expressed as an upper bound on the Lieb-Robinson commutator
\begin{align}
C_B(x,t) = \sup_{A_x} \frac{\left|\left|\left[A_x,B(t)\right]\right|\right|}{\left|\left|A_x\right|\right|},
\end{align}
where $\left|\left|O\right|\right|$ is the operator norm of the operator $O$, $A_x$ is an operator acting on site $x$ and $B(t)=\mathbb{E}_\xi B_\xi(t)$ is an operator in the Heisenberg picture that may have support over many sites. To determine $B(t)$ we require the adjoint Lindblad master equation to find the time evolution of operators. It is given by 
\begin{align}
\mathcal{L}^\dagger B(t) = i\left[H,B(t)\right] -\gamma\mathcal{D}(B)
\end{align}
where the dissipator is $D(B(t)) = \sum_jD_j(B(t)) = \sum_j\left[P_j,\left[P_j,B(t)\right]\right]$. To approximate $\left|\left|\left[A_x,B(t)\right]\right|\right|$ we take a Taylor expansion of $B(\tilde{t})$ with $\tilde{t}=t+\epsilon$. Up to order $\epsilon^2$, we have
\begin{align}
\left|\left|\left[A_x,B(\tilde{t})\right]\right|\right| = ||\left[A_x,B(t)\right]+\epsilon\left[A_x,\partial_t B(t)\right]\\\nonumber
+\frac{\epsilon^2}{2}\left[A_x,\partial_t^2 B(t)\right]+\mathcal{O}(\epsilon^3)||.
\end{align}

Now, we expand each term individually. First, the $\mathcal{O}(\epsilon)$ term is given simply by the adjoint master equation, i.e.,
\begin{align}
\epsilon\left[A_x,\partial_t B(t)\right]&=\epsilon\left[A_x,i\left[H,B\right]-\gamma\mathcal{D}\right]\\
&= i\epsilon\left[A_x,\left[H,B\right]\right] -\epsilon\gamma\left[A_x,\mathcal{D}\right].
\end{align}
The $\mathcal{O}(\epsilon^2)$ term requires taking the time derivative of the time-dependent terms of the adjoint master equation, as follows:
\begin{align}
\frac{\epsilon^2}{2}\left[A_x,\partial_t^2 B(t)\right] &= \frac{\epsilon^2}{2}\left[A_x,\partial_t (i\left[H,B\right]-\gamma\mathcal{D})\right]\\
&=\frac{i\epsilon^2}{2}\left[A_x,\left[H,\partial_t B\right]\right] -\frac{\epsilon^2\gamma}{2}\left[A_x,\partial_t\mathcal{D}\right]\\
&=\frac{-\epsilon^2}{2}\left[A_x,\left[H,\left[H,B\right]\right]\right]\nonumber\\
&\hspace{2cm}- \frac{i\epsilon^2\gamma}{2}\left[A_x,\left[H,\mathcal{D}\right]\right]\nonumber\\
&\hspace{2.5cm}-\frac{\epsilon^2\gamma}{2}\left[A_x,\partial_t\mathcal{D}\right].
\end{align} 
Assuming that $\sum_jP_jB\leq pB$, the final term is given by
\begin{align}
[A_x,\partial_t\mathcal{D}] = 2ip&\left[A_x,\left[H,B\right]\right]-2\gamma p\left[A_x,\mathcal{D}\right]\nonumber\\
&-2[A_x,\sum_j P_j\partial_t B(t) P_j].
\end{align}
Only keeping $\mathcal{O}(\epsilon^2)$ terms that also appear at $\mathcal{O}(\epsilon)$ and using the triangle inequality results in the following bound:
\begin{align}
\left|\left|\left[A_x,B(\tilde{t})\right]\right|\right|&\leq (1-\epsilon\gamma p)\left|\left|\left[A_x,B(t)+i\epsilon\left[H,B(t)\right]\right]\right|\right| \nonumber\\
&+ \epsilon\gamma\left|\left|\left[A_x,pB+(\epsilon\gamma p-1)\mathcal{D}\right]\right|\right|+\mathcal{O}(\epsilon^2).
\end{align}

We now bound each term individually. Using the unitary equivalence of the operator norm and by repeated Taylor expansions, the first term is bounded by
\begin{align}
\left|\left|\left[A_x,B(t)+i\epsilon\left[H,B(t)\right]\right]\right|\right|&\leq \left|\left|\left[A_x,B(t)\right]\right|\right|\nonumber\\
&+\epsilon\left|\left|\left[\left[H,A_x\right],B(t)\right]\right|\right|.
\end{align}
The second term is
\begin{align}
\left|\left|\left[A_x,pB+(\epsilon\gamma p-1)\mathcal{D}\right]\right|\right|&\leq p\left|\left|\left[A_x,B(t)\right]\right|\right|\nonumber\\
&+(\epsilon\gamma p -1)\left|\left|\left[A_x,\mathcal{D}\right]\right|\right|.
\end{align}
Again, keeping only terms that already appear at $\mathcal{O}(\epsilon)$ the bound is reduced to
\begin{align}
\left|\left|\left[A_x,B(\tilde{t})\right]\right|\right|\leq &(1-2\epsilon\gamma p+2\epsilon^2\gamma^2p^2)\left|\left|\left[A_x,B(t)\right]\right|\right|\nonumber\\
&+ \epsilon(1-\epsilon\gamma p)\left|\left|\left[\left[H,A_x\right],B(t)\right]\right|\right|+\mathcal{O}(\epsilon^2).
\end{align}

It follows that the time derivative of the Lieb-Robinson correlator is bound by
\begin{align}
\partial_t C_B(x,t)\leq -2\gamma pC_B(x,t)+\sup_{A_x}\frac{\left|\left|\left[\left[H,A_x\right],B(t)\right]\right|\right|}{\left|\left|A_x\right|\right|}.
\end{align}
In order to solve this, we must expand the double commutator on the right. First note that, for our fermionic system with nearest-neighbor interactions, $\left[H,A_x\right]=2\left|\left|H\right|\right|\left|\left|A_x\right|\right|V$ with $V=u_{x-1}n_{x-1}n_x+u_xn_xn_{x-1}$ and $|u_y|\leq\left|\left|V\right|\right|\leq 1$. The operator $V$ will contribute both diagonal and off-diagonal components and can be bound by $V\leq 2\mathbb{I}+R$, where $R_{j,k}=\delta_{j,k+1}+\delta_{j+1,k}$. We may also express components $C_B(x,t)$ as a vector $C_B(t)=[C_B(1,t),\dots,C_B(N,t)]$, resulting in the bound 
\begin{align}
\partial_t C_B(t)\leq ((-2\gamma p+4\left|\left|H\right|\right|)\mathbb{I}+2\left|\left|H\right|\right|R)C_B(t),
\end{align}
where the inequality is evaluated component-wise. This bound emits the final solution
\begin{align}\label{eq:lrb_pm_supp}
C_B(x,t)\leq e^{(-2\gamma p+4\left|\left|H\right|\right|)t}\sum_j^N\left(e^{2\left|\left|H\right|\right|Rt}\right)_{x,j}C_B(j,0).
\end{align}

\bibliography{references}

\end{document}